\documentclass{article}
\usepackage[english]{babel}
\usepackage[utf8]{inputenc}
\usepackage{johd}
\usepackage{amsfonts}
\usepackage{caption}
\usepackage{subcaption}
\usepackage{multicol}
\usepackage{graphicx}

\title{Quantitative Comparison of Nearest Neighbor Search Algorithms}

\author{\small Hanitriniala Malalatiana 
Rakotondrasoa,${}^{abd}$\footnote{Corresponding author: malalarhm at aims.ac.za} Martin Bucher${}^{abcd}$ and Ilya Sinayskiy${}^{ad}$\\
      \small
  $^{a}$School of Chemistry and Physics, University of KwaZulu-Natal, Durban, South Africa \\
      \small
        $^{b}$School of Data Science and Computational Thinking, University of Stellenbosch, Stellenbosch, 7600, South Africa\\
      \small
        $^{c}$Laboratoire Astroparticule et Cosmologie (APC), Universit\'e Paris Cit\'e/CNRS, F-75013 Paris, France\\
      \small
        $^{d}$National Institute for Theoretical and Computational Sciences (NITheCS), South Africa
}

\date{} 

\begin{document}

\maketitle

\begin{abstract} 
\noindent 
We compare the performance of three 
nearest neighbor search algorithms: the Orchard, ball tree, and VP-tree
algorithms. 
These algorithms are commonly used for nearest-neighbor searches and are known for their efficiency in large datasets. We analyze the fraction of distances computed in relation to the size of the dataset and its dimension. 
For each algorithm we 
derive a fitting function for the efficiency as a function to set size and dimension. 
The article aims to provide a comprehensive analysis of the performance of these algorithms and help researchers and practitioners choose the best algorithm for their specific application.  \end{abstract}

\noindent\\


\begin{multicols}{2}
\section{Introduction}

Nearest neighbor searches arise in a wide variety of applications such as image retrieval, recommender systems, anomaly detection, and genomics (\cite{wang2010locality}, \cite{koren2009matrix}, \cite{chandola2009anomaly}, \cite{parry2010k}).
In a nearest neighbor search problem, one is given a finite set of points $S$ that is the subset of a larger
usually infinite set of points $X$ on which a metric $d(~\cdot ~, ~\cdot ~)$ is defined obeying the usual axioms
for a metric space. Given a query point $q\in X,$ the nearest neighbor search recovers a point $s\in S$ (not 
necessarily unique) such that $d(q,s)$ is minimized \citep{Kaufmann}. 
Let us call the result $N\!N(q, S)$. In the case where the result is not unique, this may not be a function as we can allow $N\!N$ 
to return any of the minima rather than an exhaustive list.
Special techniques exist for the case where $X$ is Euclidean  
and of low dimension. For $X=E^1$ the point can be ordered and well-known binary search techniques
can be applied \citep{Knuth1973}. For the $E^2$ special techniques exist that divide the plane into zones, and similar techniques
can be generalized to $E^k$ where $k$ is small (\cite{aurenhammer1991voronoi}, \cite{klee1979complexity}). However, in many cases, $X$ cannot be embedded into a Euclidean
space, and even when that may be possible, the structure of the set $S$ is such that its dimension is fractional
or much lower than that of the space into which it can be embedded so that algorithms applicable to a general
metric space with no special structure beyond the usual axioms are most useful. 

A brute force solution to the nearest neighbor problem is trivial. If $N=\vert S\vert$, calculating 
$d(q,s)$ for each $s\in S$ requires $O(N)$ effort and is guaranteed to return the correct solution with probability one.
The objective is to do better: to find the solution by making as few comparisons as possible. This often involves 
arranging the elements $S$ into a suitable pre-computed data structure, which may be exploited 
to find $N\!N(q,s)$ with a number of comparisons significantly smaller than $N.$ The property of
the metric space that allows unnecessary comparisons to be avoided is the triangle inequality. Considering
the trivial case when $X=E^1$ sets the lower bound on the possible efficiency of the algorithm to be 
$O(\log _2(N))$ which results from a balanced binary tree. 

Here we explore by numerical experiment the efficiency of three 
nearest neighbor search
algorithms applicable 
to the case where $X$ is a metric space with no special structure: the Orchard algorithm, the ball tree algorithm, and the VP (vantage point)-tree algorithms. We find a novel fitting functions for the efficiency of the algorithm compared to the brute force algorithm described above.

\section{Algorithms}

\subsection{Orchard Algorithm}

 The Orchard algorithm, invented by Michael T. Orchard \citep{orchard1991fast}, exploits the triangle inequality to reduce the distance calculated in searching for the nearest neighbor. Orchard algorithm is simple and fast, particularly in high dimension ($\textrm{dim}~X< 64$, for example) \citep{Zatloukal}. 
 
 The Orchard algorithm relies on a pre-computed array of size $\mathcal{O}(N^2)$ where $S_{pointer}[i,j]$ (where $i=0 \dots (N-1)$ and $j=0 \dots (N-2)$) points to the $j$-th closest point to $S_i$. In other words, $S_{pointer}[i,:~ ]$ includes all the $(N-1)$ points not equal to $S_i$ arranged according to increasing distance from $S_i$. Setting up this table requires $\mathcal{O}(N^2)$ operations and $\mathcal{O}(N^2)$ storage space is required.

\begin{figure}[H]
\begin{center}
\includegraphics[width=0.4\textwidth]{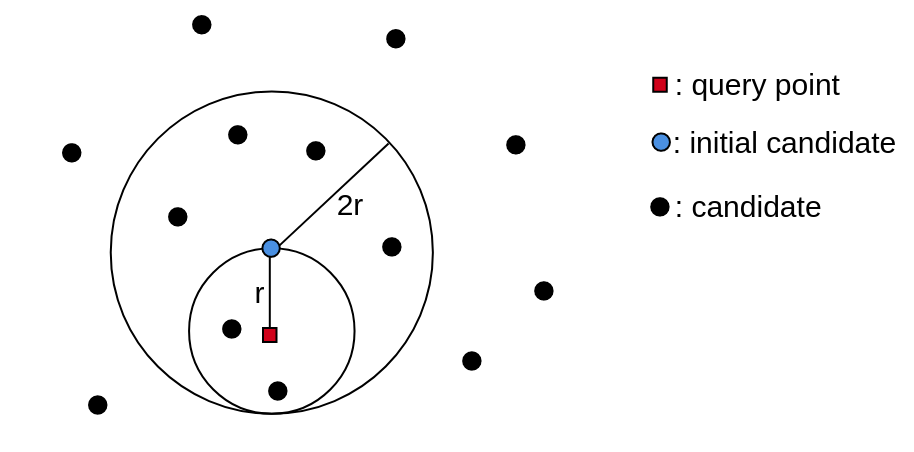}
\end{center}
\caption{The Orchard algorithm 
limits the number of points that need to be tested by exploiting the triangle inequality. If the current closest point (in the center
of the large circle) is a distance $r$ from the query point, any point closer to the query point must lie with a radius $2r$ of the current closest point. The property provide a stopping rule for the algorithm.}
\label{Orchard}
\end{figure}

\figurename{ \ref{Orchard}} illustrates the idea behind the algorithm. The idea is to eliminate all candidates outside the circle centered on the initial candidate. Given a query (shown in red), it begins by selecting a point at random that is supposed to be the closest candidate. Knowing the distance $r$ between the query and the chosen nearest candidate allows one to skip computing the distance between the query and the other points outside the circle with radius $2r$ centered on the chosen nearest candidate.

Suppose we have access to the distances $d(q, S_i)$ between a query point and a selected candidate, as well as the distances $d(S_j, S_i)$ between the selected candidate and other candidates. In the search process, an optimization technique arises when $d(S_j, S_i)$ exceeds twice the distance $d(q, S_i)$. In such cases, there is no need to compute the distance $d(q, S_j)$ between the query point and the other candidate $S_j$. This property is derived from the triangle inequality.

The search algorithm begins by randomly selecting a point $S_i$ from a given set of points $S$. Then it calculates the distance $d(S_i, q)$ between this selected point and a query point $q \in X$. Next, the algorithm proceeds to iterate through each point $S_j$ in $S_{pointer}[i,:]$. If $d(S_j, S_i) < 2d(q, S_i)$, the algorithm calculates the distance between the current point $S_j$ and the query point $q$. If $d(S_j, q) < d(S_i, q)$, the algorithm updates the index from $i$ to $j$, indicating that the current point $S_j$ is more relevant. The algorithm continues this iterative process within the updated subset $S_{pointer}[i,:]$ until relevant points have been considered.

\subsection{Ball Tree Algorithm}

The ball tree algorithm organizes the points of $S$
into a balanced binary tree, so that each element $s$ of $S$
is assigned to a unique node of the tree \citep{uhlmann1991satisfying}.
Each node $\mathcal{N}$ additionally includes a real number $\mu \ge 0$ (i.e., the ball radius) 
such that $\mu =\textrm{max}_{l\in \mathcal{L}}d(l,s)$
and $\textrm{min}_{r\in \mathcal{R}}d(r,s)\ge \mu $ 
as well as pointers (possibly NULL) to the 
left and right child nodes.
Here $\mathcal{L}$ is the set of 
all descendant points of the left branch
of $\mathcal{N},$ and $\mathcal{R}$ is the set of all the descendant points  of the right
branch of $\mathcal{N},$ and the cut is chosen such that $\vert \mathcal{L}\vert =\vert \mathcal{R}\vert $ or
$\vert \mathcal{L}\vert =\vert \mathcal{R}\vert +1.$ Leaf nodes have no children,
and node having only a left child are allowed. 

The process of
setting up the ball tree data structure
may be described recursively.
A node $\mathcal{N}$ is generated
from a set of remaining points
$S'$ by selecting and removing 
a random point $s.$
Define $\bar S'=S'-\{ s \} .$
$\bar S'$ is partitioned into two disjoint subsets
$\mathcal{L}$ and $\mathcal{R}$ as described above. 
The left child node $\mathcal{N}_\mathcal{L}$ is generated 
from $\mathcal{L}$ by applying the same procedure described above, and similarly the right 
child node $\mathcal{N}_\mathcal{R}$ is generated from $\mathcal{R}$ in the same way.

The storage space of the ball tree structure is $O(N),$
and the computation required to set up this structure is
$O(N\log N).$

\subsection{Vantage Point Tree}

In the ball tree algorithm, in the process of 
setting up the ball tree data structure, after
$S'$ was partitioned into two disjoint subsets
$\mathcal{L}$ and $\mathcal{R},$ the points $l$ and $r$ where randomly
drawn from $\mathcal{L}$ and $\mathcal{R}$ to continue the 
tree downward. While these random choices provide
a fast way to set up the ball tree structure, these
choices are not necessarily the most efficient 
for carrying out a subsequent query search using
the ball tree structure. The vantage point algorithm
rather than making a random choice attempts to
optimize the choice of $l$ and $r$ to improve
the efficiency of the subsequent query searches.

\cite{yianilos1993data} has shown with an example in a $[0,1]$ bounded metric space that points near the space's corners can provide the best vantage point. 
In the ball tree algorithm, a vantage point $v\in S'$ was chosen at random.
But some vantage point choices are better than others. A good vantage point
has few points near the boundary at $r=\mu (v)$, which is the middle value of distances from $v$ to each point in the random subsample $S'-\{v\}$, dividing the set into equal
halves. The vantage point is optimized so that the second moment 
\begin{equation}
\left\langle \left(
d(p,v)-\mu (v) 
\right) ^2 
\right\rangle _{p\in S'-\{ v \} }
\end{equation}
is maximized. This choice 
minimizes the probability of encountering the situation illustrated
in Fig.~\ref{search}(c) where the search path bifurcates, leading to more distance evaluations.

In practice only a random subset of candidate vantage points are explored, 
and the second moment above is approximated using only a random subsample
of $S'-\{ v \} .$ A vantage points needs to be good but not necessarily
the very best.  The vantage point method leads to a tree in which less 
branches need to be explored, leading to behavior more closely
ressembling $O(N\log (N)).$



\subsection{Nearest Neighbor Query Using Ball Tree Structure}

As the query algorithm is common to both the ball tree and VP algorithms, we discuss here  the query search based
on this pre-computed tree data structure. 
Both these algorithms attempt to apply divide-and-conquer in analogy to a binary search tree for totally ordered data. 
This analogy however is imperfect. In the binary search, one descends the tree only once without ever having to backtrack, 
whereas here in general one must backtrack at least sometimes, and in the worst-case scenario the entire tree must be traversed. 
The ball tree algorithm provides the capability to find not only one nearest neighbor for a given query point $q$ but also multiple 
nearest neighbors within a specified radius $\epsilon$ the from $q$ defined at the start of the search. On the other hand, if we only require the single nearest neighbor, we 
can update the value of $\epsilon$ 
with each minimum distance encountered during the search. This strategy allows us to continually refine the search space and 
focus on identifying the closest point to $q$. 
                    
\begin{figure}[H]
\begin{center}
\includegraphics[width=0.4\textwidth]{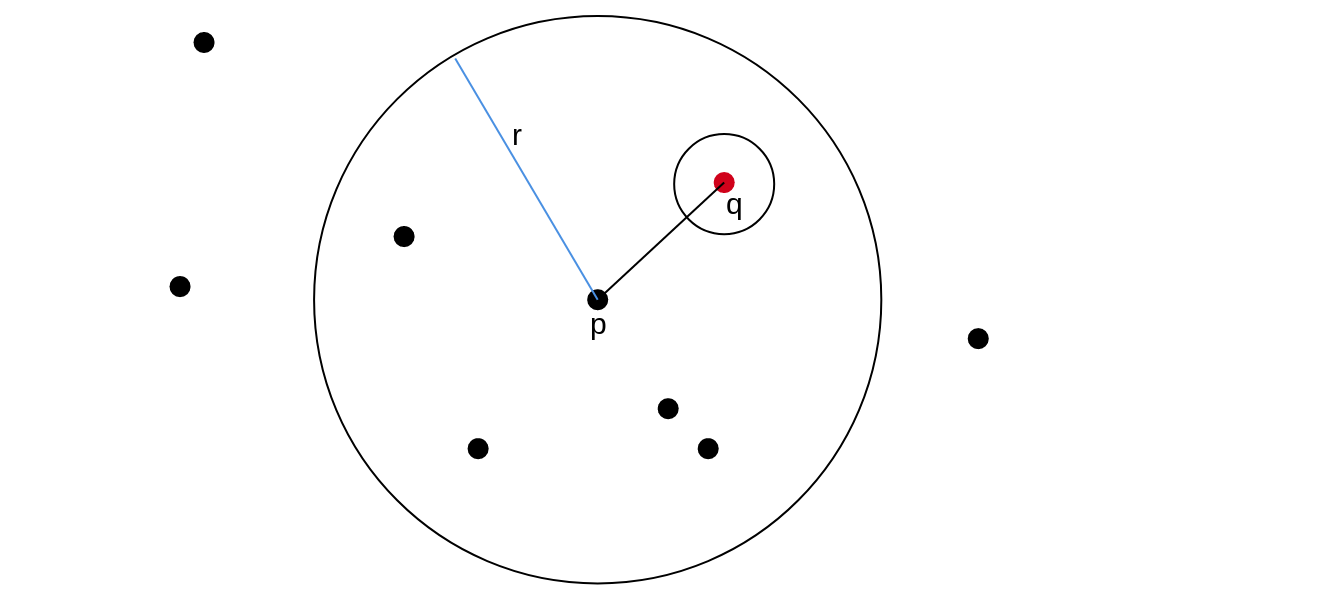} \\
(a)\\
\includegraphics[width=0.4\textwidth]{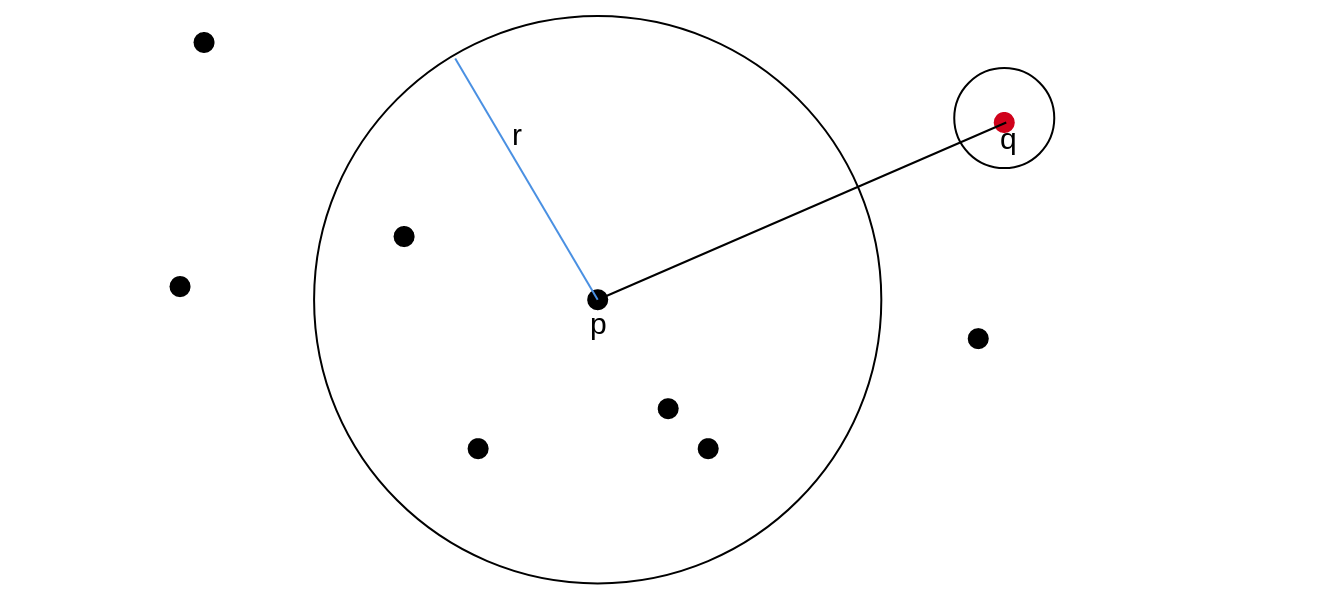}\\
(b)\\
\includegraphics[width=0.4\textwidth]{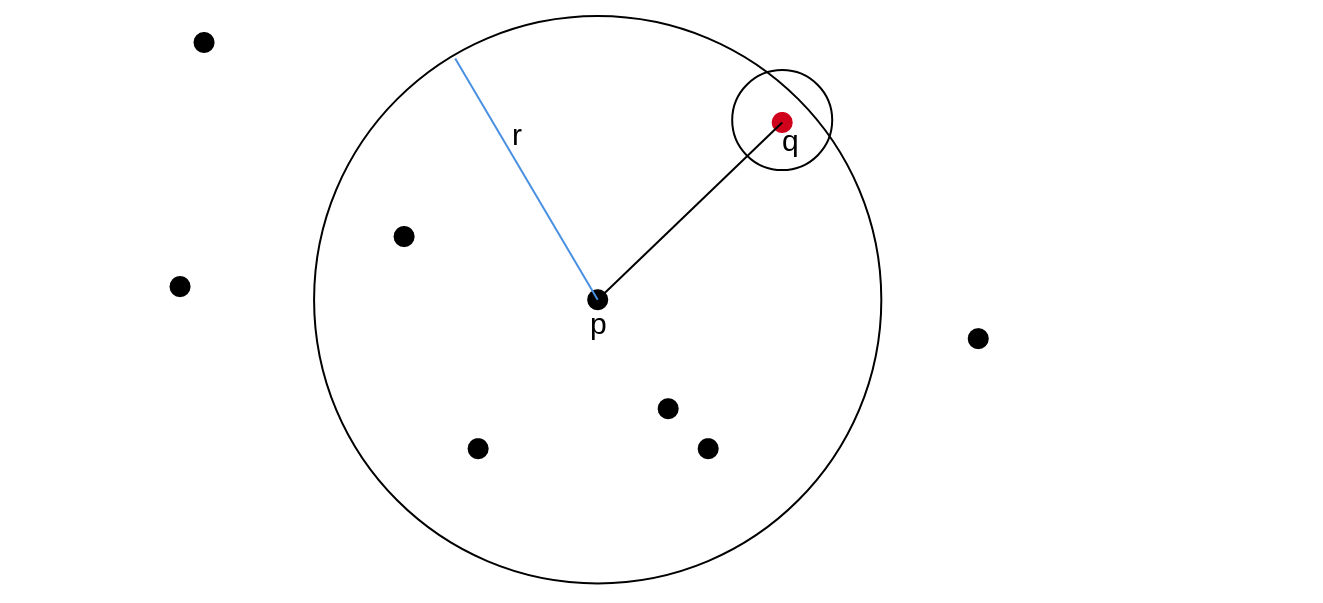}\\
(c)\\
\end{center}
\caption{The small circle is a ball of radius $\epsilon $
about the query point where $\epsilon $ indicate the distance to the closest point found 
so far. In case (a) only the left subtree is explored and in case (b) only the right tree
is explored. Case (c) however is the worst case scenario because in this case both branches must be explored.}
\label{search}
\end{figure}

When searching for the closest point to a given point, we start at the root node and work our way down the tree, comparing the distance between the point we are looking for and the ball of each node. If the distance between the point we are looking for and the ball is greater than the distance between the point we are looking for and the nearest point already found, we can ignore that branch of the tree altogether, as it won't contain the nearest point.
Given a query $q$ and a query radius $\epsilon$. One wishes to find all objects $o$ with a distance less than $\epsilon$, i.e. $d(q,o)<\epsilon$. When visiting a node $S_i$ with a ball radius $r$, one must decide whether to visit the left or right child of that node. If $d(S_i, q) > r + \epsilon$, the search in the left child may be excluded, and if $d(S_i, q) < r - \epsilon$, the search in the right child may be excluded. However, we cannot avoid the search on the left and right if $r-\epsilon<d(S_i,q)<r+\epsilon$ \citep{Kaufmann}. 
Figure \ref{search} shows the three cases.

\subsection{Worst Case Scenario}

The triangle inequality alone offers a powerful tool to avoid unnecessary comparisons but it is not in all cases possible to improve on the brute force algorithm. Suppose $S$ is such that $d(S_i,S_j)=D$ for all $i,j=0 \dots (N-1)$ and moreover the query point $q$ is such that $d(S_i,q)=D$ as well. Such a set can be embedded in $N$-dimensional Euclidean space. In this case, none of the algorithms above improve on the brute force algorithm. Perturbing each of the points by a small amount so that the distances fall in the interval $[D-\epsilon, D+\epsilon]$ where $\epsilon$ is sufficiently small in less contrived but does not change the situation.

The  worst case scenario just described is not as contrived  or exceptional as might seem at first sight. Let us take $N$ points distributed according to a Gaussian in $D$-dimensional Euclidean space.
\begin{equation}
    p(\mathbf{x})=\left(\frac{2\pi}{D}\right)^{D/2}\exp\left[-\frac{1}{2}D
    \mathbf{x}^T\mathbf{x}
    \right]
\end{equation}
As $D \to \infty$, all the points very nearly lie on the unit sphere. The fluctuation in the radius is of order $\mathcal{O}(1/\sqrt{D})$. The distances between pairs of points are similar. This is known as the ``curse of dimensionality.''

\section{Benchmark Description}

In order to compare and benchmark nearest neighbor algorithms, we define two-parameter family test cases, which we hope are representative of these algorithms in real word applications. The two parameters are the number of points $N$ and dimension $D$. The $N$ points are independently drawn from a Gaussian distribution 
\begin{equation}
    p(\mathbf{x}) \sim \exp\left[-\frac{1}{2}\mathbf{x}^T\mathbf{x}\right]
\end{equation}
where $\mathbf{x} \in \mathbb{R}^D$. Distances are computed using the Euclidean metric. One may consider generalizing by replacing $\mathbf{x}^T\mathbf{x}$ with 
$\mathbf{x}^T\mathbf{C}^{-1}\mathbf{x}$ where the covariance matrix $\mathbf{C}$ is positive definite. The overall scale of $\mathbf{C}$ is irrelevant: $\mathbf{C} \mapsto \lambda \mathbf{C}, \lambda >0$ does not produce a set of points with different behavior. However, the condition number of $\mathbf{C}$, $K=\lambda_{max}/\lambda_{min}$ can produce different behavior where $K\gg 1$, for example when the effective dimension depends on $N$. The Gaussian form however probably is not so material. However, the assumption of lack of clustering or Poissonian statistics could be important. One imagines the distribution of points with hierarchical clusters, where behavior might differ substantially. We did not investigate these issues because we could not find a family of test cases that would be representative of all the types of behavior that could be encountered.

\section{Quantitative Comparisons}

The metric to compare the three algorithms among themselves and with the brute force algorithm
is $f$ the ratio of the number of comparisons made compared to the number of comparisons required with the brute force algorithm, which is equal to $N$. This fraction is a random variable, as the test problem was defined above as the outcome of a random process. Consequently, we compute an average over many realizations.

\begin{figure}[H]
    \centering
    \includegraphics[width=0.35\textwidth]{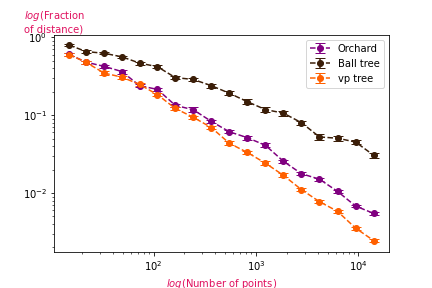}    
    \caption{Orchard, ball tree, and VP tree algorithm 
    efficiency for different numbers of points at $d=3$}
    \label{companpoint}
\end{figure}

\figurename{ \ref{companpoint}} shows $f$ as the function of the number of points for $d=3.$
With this metric the Orchard algorithm outperforms 
the Ball tree, although the increased storage requirement and pre-computation for the Orchard algorithm should be considered as well. The
VP-tree algorithm however outperforms both the ball
tree and Orchard algorithms. In all cases, as the number of the points increases, the mean fraction of distances decreases. Additionally, the utilization of vantage point selection proves to enhance search performance compared to random selection. 
On the log-log plot one observes linear behavior indicating that a power law for$f$ gives a good fit.

\begin{figure}[H]
    \centering
    \includegraphics[width=0.35\textwidth]{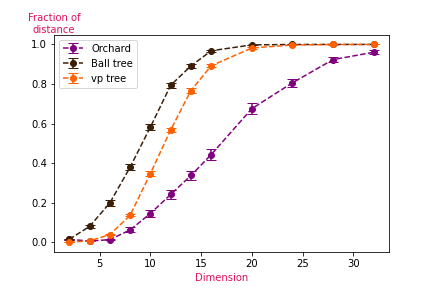} \caption{Orchard, ball tree, and VP tree algorithm 
    efficiency with $9000$ points as a function of dimension.}
    \label{compandim}
\end{figure}

\begin{figure}[H]
\begin{center}
\includegraphics[width=0.4\textwidth]{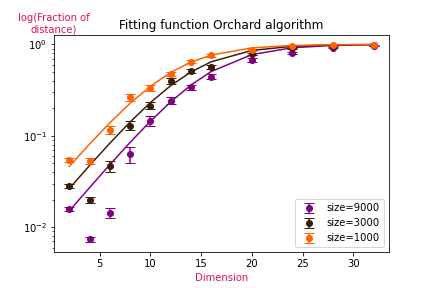}
\includegraphics[width=0.4\textwidth]{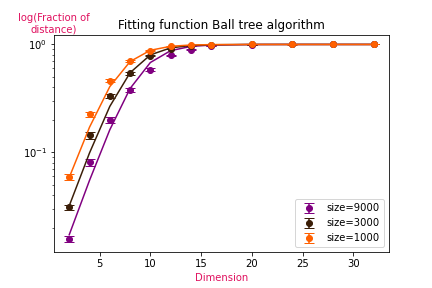}
\includegraphics[width=0.4\textwidth]{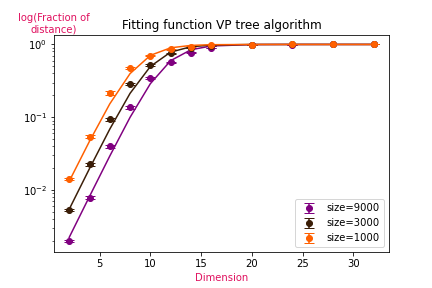}
\end{center}
\caption{Comparison of fitting function to the data for the Orchard, ball tree, and VP tree algorithms}
\label{fitting}
\end{figure}

\figurename{ \ref{compandim}} explores
datasets of various sizes and dimensions,  
with $\vert S\vert =$
1000, 3000, and 9000 and dimension $d$ from 2 to 32. 

These results show that the Orchard algorithm outperforms both the Ball tree and VP-tree 
algorithms in terms of computational efficiency. The Orchard algorithm is observed 
to compute fewer distances compared to the other two algorithms. However, despite 
this advantage, it is the VP-tree algorithm that emerges as the more favorable 
choice, primarily due to its minimal memory requirements.

The fraction of distances evaluated is accurately 
approximated by the sigmoid function
\begin{equation}
f(D,N)= \frac{1}{1+ \exp [-\beta (D-\alpha \log(N)]}.    
\end{equation}
For each algorithm the parameters
$\alpha$ and $\beta$ are fitted to the data
using generalized least squares. 
The best fit parameters are: $\alpha=1.75$ and $\beta=0.3$ for the Orchard algorithm;
$\alpha=0.96$ and $\beta=0.6$
for the ball tree algorithm; and $\alpha=1.25$ and $\beta=0.65$
for the VP tree algorithm, as shown in
\figurename{ \ref{fitting}}.

\section{Summary}

The emprical fitting function found here provides a 
good predictor of the performance of the three algoritms studied.
 The Orchard algorithm performs well for smaller data sets
 whereas the VP-tree algorithm is more efficient for larger 
 data sets. The ball tree algorithm is the least efficient 
 of the three algorithms. For a real application, it may not always
 be obvious what value to use for the dimension $D$ as the effective dimension
 may be less than the dimension of the ambient space or it may even be fractional.
Nevertheless, we believe that the two-parameter 
benchmark used here is representative of a
broad range of applications.
 
\section*{Acknowledgements}
HMR thanks NITheCS for support during her Master's studies at UKZN. The three authors thank the NITheCS program on Bioinformatics, Genomics, and Advanced Medicine for its support.

\bibliographystyle{johd}
\bibliography{bib}
\end{multicols}

\end{document}